\documentclass[
 reprint,
 amsmath,amssymb,
 aps,
 prl,
]{revtex4-1}

\usepackage{graphicx}
\usepackage{dcolumn}
\usepackage{bm}
\usepackage{color}
\usepackage{bbold}
\usepackage{enumerate}

\newcolumntype{L}[1]{>{\raggedright\arraybackslash}p{#1}}
\newcolumntype{C}[1]{>{\centering\arraybackslash}p{#1}}
\newcolumntype{R}[1]{>{\raggedleft\arraybackslash}p{#1}}

\begin{document}

\preprint{preprint: [prb]}

\title{Interplay of resonant states and Landau levels in functionalized graphene}

\author{Jeongsu Lee}
\email{jeongsu.lee@physik.uni-regensburg.de}
\affiliation{
Institute for Theoretical Physics, University of Regensburg, 93040 Regensburg, Germany
}
\author{Denis Kochan}
\email{denis.kochan@physik.uni-regensburg.de}
\affiliation{
Institute for Theoretical Physics, University of Regensburg, 93040 Regensburg, Germany
}
\author{Jaroslav Fabian}
\email{jaroslav.fabian@physik.uni-regensburg.de}
\affiliation{
Institute for Theoretical Physics, University of Regensburg, 93040 Regensburg, Germany
}

\date{\today}

\begin{abstract}
Adsorbates can drastically alter physical properties of graphene. Particularly important are
adatoms and admolecules that induce resonances at the Dirac point. Such resonances limit
electron mobilities and spin relaxation times. We present a systematic tight-binding as
well as analytical modeling to investigate the properties of resonant states in the presence of 
a quantizing magnetic field. Landau levels are strongly influenced by the resonances, especially 
close to the Dirac point. Here the cyclotron motion of electrons around a defect leads to  the
formation of circulating local currents which are manifested by the appearance of side peaks
around the zero-energy Landau level. Our study is based on realistic parameters for H, F, and
Cu adatoms, each exhibiting distinct spectral features in the magnetic field. We also show
that by observing a local density of states around an adatom in the presence of Landau levels
useful microscopic model parameters can be extracted. 
\end{abstract}
\maketitle

\section{Introduction}

Graphene on a substrate is essentially a surface which is intrinsically susceptible to contamination with adsorbates---atoms and molecules---that can significantly affect the electronic properties \cite{CastroNeto2009RMP, Wehling2009, Wehling2010, Loktev2007PRB, Loktev2010PRB, Han2014NNano}.  Adsorbates can generate giant spin-orbit coupling 
\cite{CastroNeto2009PRL, Gmitra2013PRL, Kochan2014PRL, Zollner2016PRB, Frank2017PRB, Kochan2017PRB, Pachoud2014PRB, Weeks2011PRX, Mertig2010PRB, Zhou2010Carbon, Avsar2015TDMatt, Balakrishnan2013NPhys}
and even induce local magnetic moments, as predicted theoretically \cite{Yazyev2007PRB} and confirmed in experiments 
\cite{Hong2011PRB, McCreary2012PRL, Tang2105SR, Gonzalez2016Science}.
Relevant for transport, strong covalent bonds between adsorbates and host carbon atoms can be manifested as infinite point-like scatterers. Dirac electrons bouncing off such centers experience resonant scattering which can be seen as prominent peaks in the density of states or scattering probabilities formed at energies close to the Dirac point. At sufficiently large adsorbate densities, these scattering centers can dominate over other scattering mechanisms and limit the electron mobilities \cite{Geim2010NLett, Ozyilmaz2011ACSNano, Zhu2015PRB, Kawakami2018PRL}.
Apart from transport, adsorbate induced changes in the local electronic structure can be also read from scanning probe experiments 
\cite{Andrei2012RPP}.

In the presence of an external magnetic field electrons in graphene
exhibit cyclotron motion which is quantized in the Landau levels whose structure in graphene is qualitatively different from that in 2D semiconductors due to the massless nature of the electronic states \cite{CastroNeto2009RMP}.
While Landau level physics has been investigated mainly in clean samples, it is interesting to study the interplay between resonant scatterers and cyclotron dynamics. Electrons experiencing resonant scattering  ``stay'' longer at the resonant center, forming quasi-bound states that extend spatially farther than exponentially localized bound states. Landau levels can then form directly at the scattering centers, 
affecting the local electronic structure at resonances. This was already pointed out in the analytical treatment of Silvestrov \cite{Silvestrov2014PRB}, who showed that 
the impurity level hybridizes with one of the Landau Level (LL) states and forms two split states, and presented analytical solutions for the resonant-impurity-induced states 
in the limit of a large (compare to the lattice period) magnetic length.

Here we analyze this problem from two different angles: numerical tight-binding and
analytical Green's function, while {\it using realistic models of adatom scatterers}, parametrized from density functional calculations. The three adatoms we chose are
all important and they represent different regimes of resonant scattering. Hydrogen is the strongest resonant scatterer, akin to 
vacancies in graphene, as it generates a resonant peak very close to the 
Dirac point. Hydrogenation of bilayer graphene has recently been shown to 
give strong resonant peaks in AB stacked structures 
\cite{Kawakami2018PRL}
, while earlier results on spin transport seems to depend
on the hydrogenation mechanism \cite{Balakrishnan2013NPhys, McCreary2012PRL, Kaverzin2015PRB, Wojtaszek2011JAP}
Copper is a common adatom in epitaxial graphene, since copper substrates are used in the CVD growth. As an adatom
Cu also induces resonant scattering, albeit with a less pronounced peak in the density of states off the Dirac point \cite{Frank2017PRB, Irmer2018PRB}.
We view this case as intermediate resonance scattering . Finally, fluorine adatoms form what we call marginal resonant scatterers 
\cite{Irmer2015PRB}.
Shown by density functional calculations and subsequent tight-binding parametrization, F adatoms induce a very broad "resonant" feature at around 200 meV below the Dirac point. On the other hand, some experiments on fluorinated graphene \cite{Hong2011PRB}
appear to be consistent with a point-like resonant scattering model \cite{Zhu2015PRB}.  

We study the three adatoms in graphene in the presence of quantizing magnetic fields. Analytically, we present the Green's functions 
which can be combined with any scatterer to provide reliable scattering
amplitudes as a function of the magnetic field. Numerically, we study
large-scale systems to find local changes of the quantizing fields. 
Both approaches give the same electronic densities of states. The 
three adatoms that we study give qualitatively different responses
in the presence of a magnetic field.
In the limiting case of electron hole symmetry, 
a strong resonant scatterer can be approximated as a vacancy,
which forms two bound states interacting with the zeroth Landau level (LL$_0$) near the Dirac point. For an intermediate resonant scatterer, such as Cu, on the other hand,
only one bound state is pronounced, and its energy is relatively
insensitive to the external magnetic field. Finally, 
the marginally resonant scatterer (here F)  exhibits 
a large spectral width (in density of states or transmission amplitude) overlapping with multiple Landau levels
and forming marked side peaks at LLs.

In conventional graphene on oxide substrates such as SiO$_2$, 
scattering of electrons off of charge fluctuations is relevant 
for transport \cite{Sarma2011RMP}. But it also
happens that adatoms (in our case mainly F) can be charged, due to the electron transfer to/from graphene. It is then natural to 
ask how does the resonant scattering picture in the presence of a magnetic field change when the
scatterers themselves are charged, causing the long-range Coulomb
interaction between them and the Dirac electrons. We deal with
this problem similarly to what was investigated experimentally
and theoretically for 
vacancies 
\cite{Luican-Mayer2014PRL}.
We find that a positively charged impurity will produce one bound state in the conduction band next to LL$_0$.
We also show how the resonant states evolve with
increasing charge before and after the atomic collapse \cite{Andrei2016NPhys}.

Our specific predictions could be used in scanning tunneling experiments to obtain important electronic state characterization of resonant scatterers. We show how certain
resonant features in a magnetic field allow to extract 
on-site and hopping energies in a simple impurity model. 

\section{Model}
\subsection{Tight-binding model}
We consider a non-interacting Anderson impurity model. The unperturbed Hamiltonian of pristine graphene in a magnetic field is
\begin{equation}
H_0=- \sum_{\langle i,j\rangle} t_{ij} c_i^\dagger c_j + \text{H.c.},
\end{equation}
where $c_j^\dagger$ ($c_j$) stands for a creation (annihilation) operator of $p_z$ electron at carbon site $j$ (state ket $|c_j \rangle$), and
the nearest neighbor hopping $t_{ij}$ comprises magnetic field (Peierls substitution)
\begin{equation}
t_{ij}= t \exp \left[ i \frac{(-e)}{\hbar}\int_{\bf R_j}^{\bf R_i}{\bf A}{\cdot}d{\bf l} \right],
\end{equation}
where ${\bf A}$ is the vector potential, ${\bf R_{i,j}}$ are the coordinates of the corresponding lattice sates, 
$e$ is the positive elementary charge, and hopping parameter $t=2.6\,\text{eV}$ \cite{footnote:01}.
Throughout the paper, we assume two geometries: an infinite system---framing analytical approach, 
and a finite graphene flake \cite{footnote:02} 
with about a million carbon atoms---playground for numerical simulation.
As finite size effects get weaker with a lager system size
the flake should be much greater than the magnetic length, $l_{B}=\sqrt{\hbar/eB}$ ($\approx 5.7$~nm for $B=20$~T).

We consider three different kinds of impurities, namely, H, F, and Cu in top position.
The Hamiltonian for such an impurity is 
\begin{equation}
H_1=\varepsilon_{d} d^\dagger d + \omega_{d} (d^\dagger c_0 + c_0^\dagger d ),
\end{equation}
where $\varepsilon_d$ is the on-site energy at the impurity site,
and $\omega_d$ is the hybridization energy between the impurity and the underlying carbon atom [see Fig.~\ref{fig:01}(a)] that parametrize the first principles calculations
\cite{Rusin2011, Irmer2015PRB, Zollner2016PRB, Frank2017PRB}.
The values used in the paper are shown Table~\ref{table:01}.
Similarly as before, 
$d^\dagger$ ($d$) and $c_0^\dagger$ ($c$) create (annihilate) an electron at the impurity site and at the carbon atom below the impurity, respectively. 
We numerically calculated the local density of states (LDOS) at atomic site $i$,
$\rho_{i}(\varepsilon)=1/N\sum_n \bigl|\langle\psi_n|c_i\rangle\bigr|^2 \delta(\varepsilon-\varepsilon_n)$
where $\varepsilon_n$, $|\psi_n \rangle$, and $N$ are eigenenergies, eigenfunctions, and the order of the Hamiltonian matrix of the finite system,
employing the kernel polynomial method with the numerical tight-binding (TB) package \textsc{pybinding} \cite{Weisse2006RMP, Moldovan2016}.

\subsection{Green's function approach}

For practical purposes one can downfold impurity degrees of freedom by means of 
L\"{o}wdin's decimation procedure 
\cite{Loewdin1951}, transforming $H_1$ into the corresponding energy-dependent form 
\begin{equation}\label{Eq:LowdinFormOfPerturbaption}
H'_1(\varepsilon)= \frac{\omega_{d}^2}{\varepsilon-\varepsilon_{d}} c_0^\dagger  c_0\,.
\end{equation}
To analyze the resonant-impurity-induced bound and resonant states in the host system one needs to investigate poles of the retarded Green's resolvent in the presence of 
perturbation~\cite{Callaway1964},
$G(\varepsilon_+)=G_0(\varepsilon_+)[1-H'_1(\varepsilon_+)G_0(\varepsilon_+)]^{-1}$, where 
$\varepsilon_+=\varepsilon+i\delta$ is the energy with an infinitesimal positive imaginary part, and the unperturbed $G_0(\varepsilon_+)$ is the inverse of $\varepsilon_+-H_0$ (including proper boundary conditions).
Since in our case $H'_1(\varepsilon)$ is a localized perturbation at the lattice site that hosts $p_z$ carbon orbital $|c_0\rangle\equiv c_0^\dagger|0\rangle$ the equation 
for the bound (resonant) state energies reads $1=\langle c_0|H'_1(\varepsilon_+)|c_0\rangle\langle c_0|G_0(\varepsilon_+)|c_0\rangle$. Using Eq.~(\ref{Eq:LowdinFormOfPerturbaption}) we get
\begin{equation}\label{eq:BoundStateCondition}
\frac{\varepsilon-\varepsilon_{d}}{\omega_{d}^2}= G_0^{00}(\varepsilon, B):=\sum\limits_{n}\frac{\bigl|\langle\Psi_n|c_0\rangle\bigr|^2}{\varepsilon_{+}-\varepsilon_n},
\end{equation}
where the last equality expresses the on-site Green's function $G_0^{00}(\varepsilon, B)\equiv\langle c_0|G_0(\varepsilon_+)|c_0\rangle$ in terms of eigenenergies $\varepsilon_n$ and eigenfunctions $|\Psi_n\rangle$ of $H_0$, and the summation runs over an appropriate set of quantum numbers $n$. 

Depending on the boundary conditions the eigensystem of $H_0$ can be found analytically. 
In the case of an infinite system $|\Psi_n\rangle$'s are the well-known graphene LLs
~\cite{Zheng2002PRB, Peres2006PRB} with energies
\begin{equation}
\varepsilon_{n}=\text{sgn}(n)\sqrt{|n|} \frac{\sqrt{2}\hbar v_F}{l_{B}}\equiv \text{sgn}(n)\sqrt{|n|}\,\hbar\omega_B\,,
\label{eq:LLs}
\end{equation}
where $v_F=(3/2)t a_0/ \hbar$ is the graphene Fermi velocity and the interatomic distance in graphene $a_0\simeq1.42$~\AA.
Defining dimensionless energies $\tilde{E}=\varepsilon\bigl/\hbar\omega_B$ the on-site Green's function $G_0^{00}$ reads 
\cite{Horing2010, Rusin2011}
\begin{equation}\label{eq:OnSiteGF}
G_0^{00}(\tilde{E}, B)=-\frac{A_{\mathrm{uc}}}{\pi {l_B}^2} \frac{\tilde{E}}{2 \hbar \omega_B}\sum\limits_{n=0}^{N_B}
\frac{1}{n + 1 -\tilde{E}^2} + \frac{1}{n - \tilde{E}^2},
\end{equation}
where $A_{\mathrm{uc}}$ is the area of the graphene unit cell, and the cut-off $N_B$ is the integer part of $\pi {l_B}^2\bigl/A_{\mathrm{uc}}$. It is worth stressing that the formula for $G_0^{00}$ is valid in the energy range where the pristine graphene band structure can be approximated by the linear energy-momentum dispersion. Moreover, the chosen cut-off $N_B$ guarantees that the integral of $-\tfrac{1}{\pi}\mathrm{Im}\,G_0^{00}$ gives the correct number of states (Debye prescription) within the graphene bandwidth. For the typical magnetic fields, say, from 5 to 50\,T, the magnitudes of $N_B$ ranges roughly from 8000 to 800.
Depending on the energy, and the strength of the magnetic field the summation over $n$ in Eq.~(\ref{eq:OnSiteGF}) can be approximated in a controllable way. For example, for energies $|\tilde{E}|<1$ that are centered around the zero Landau level (LL$_0$) singularities of $G_0^{00}(\tilde{E}, B)$ stem from the $1/\tilde{E}^2$ contribution and the sum $\sum_{n=1}^{N_B} 1\bigl/(n - \tilde{E}^2)$, which can be approximated by the harmonic progression 
\begin{equation}
H_{N_B} \equiv \sum_{n=1}^{N_B} 1\bigl/n\simeq\ln{N_B}+\gamma,
\end{equation}
where $\gamma\simeq 0.57721$ is the Mascheroni constant. With this approximation, Eq.~(\ref{eq:OnSiteGF}) leads to %
\begin{equation}\label{eq:OnSiteGF2}
G_0^{00}(\tilde{E}, B)\approx-\frac{A_{\mathrm{uc}}}{\pi {l_B}^2} \frac{\tilde{E}}{2\hbar \omega_B}
\left(
-\frac{1}{\tilde{E}^2}+2H_{N_B}
\right),
\end{equation}
and the formula for the bound state energies, Eq.~(\ref{eq:BoundStateCondition}), simplifies to
\begin{equation}
\tilde{E}-\tilde{E}_{d}\approx  \frac{A_{\mathrm{uc}}}{\pi {l_B}^2} \frac{\omega_{d}^2}{( \hbar \omega_B)^2} \left[ \frac{1}{2\tilde{E}}- {H_{N_B}} \tilde{E} \right].
\label{eq:approxGR}
\end{equation}
From this equation, we arrive at an important result allows to anticipate the model tight binding parameters
from the experimentally measured bound states energies.
The total density of the states is also calculated from the unperturbed Green's function ($B\neq 0$)  
\begin{eqnarray}\label{eq:DOS}
\rho(\varepsilon, B)=-\frac{1}{\pi}\text{Im\,Tr} \left[G_0(\varepsilon) 
- \eta \frac{\partial G_0(\varepsilon)}{\partial \varepsilon} \mathcal{T}(\varepsilon)
\right]
\end{eqnarray}
where T-matrix $\mathcal{T}(\varepsilon)=H'_1(\varepsilon)[1-G_0(\varepsilon)H'_1(\varepsilon)]^{-1}$,
and we use the concentration of the impurity $\eta=10^{-5}$.

\section{Results}
\subsection{Locality of the resonance and bound states}
\begin{figure}[htp]
\includegraphics[height=2.9cm]{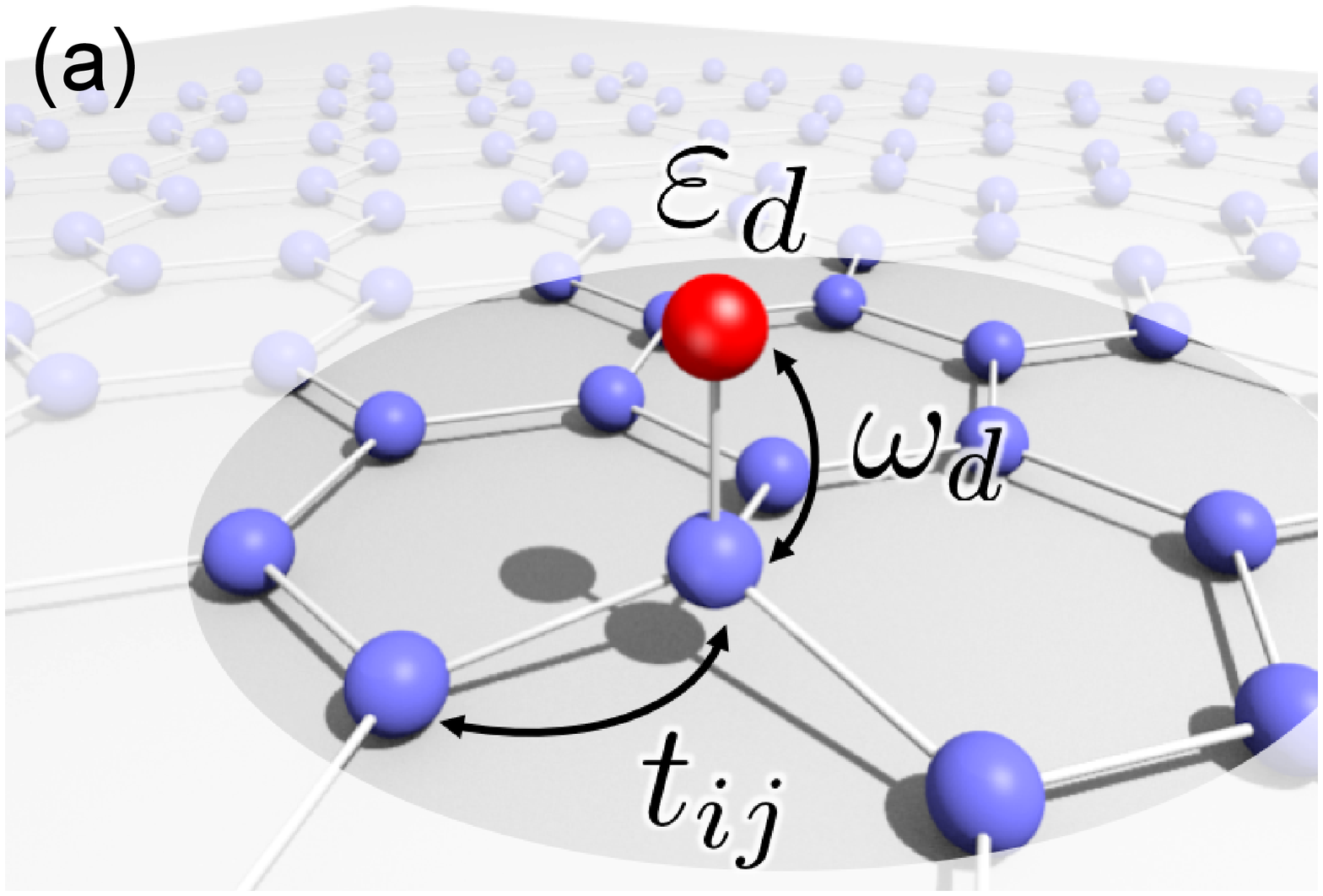}
\hspace{0.3cm}
\includegraphics[height=2.9cm]{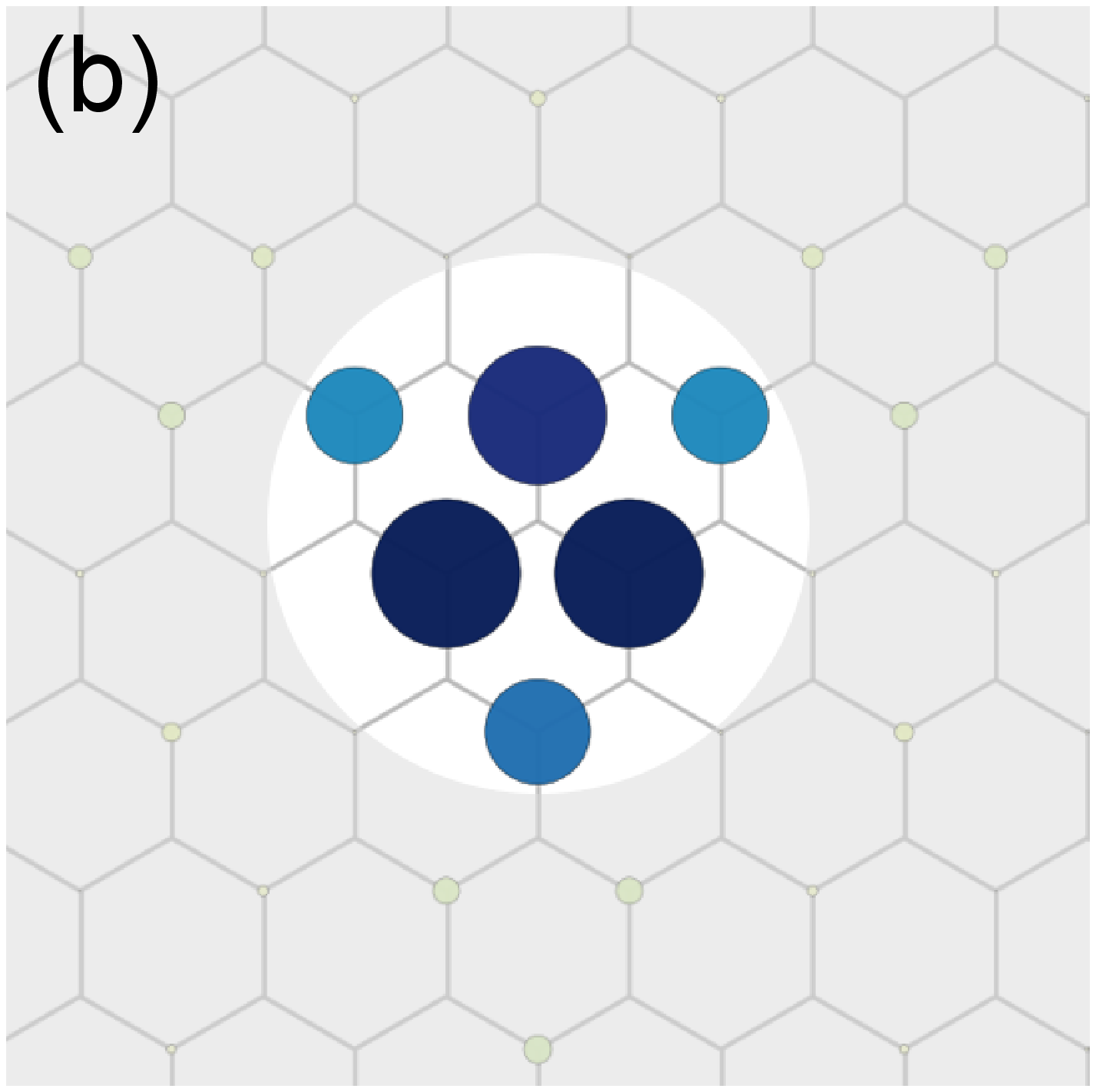}
\caption{
(a) TB parameters around the impurity.
(b) The probability density of the resonant state on the graphene sheet is shown around the impurity site (H--adatom).
Bound states ($B\neq0$) also exhibit qualitatively the same localized nature in spatial distribution.
}
\label{fig:01}
\end{figure}

One of the interesting and important aspects of resonant scattering is that the resonant and bound states form falling-off wave functions 
following the power law $\sim 1/r$ as shown in Fig.~\ref{fig:01}(b) \cite{Silvestrov2014PRB, Nanda2012NJP, Irmer2018PRB, Bundesmann2015PRB}.
For a single adatom \cite{Bundesmann2015PRB}, the probability density of the resonant state is concentrated around the impurity (or vacancy) site, mostly over the opposite sublattice.
In the presence of a magnetic field, the resonant states split into two bound states as discussed below.

\subsection{Three patterns}
\begin{figure}[tbp]
\includegraphics[width=\columnwidth]{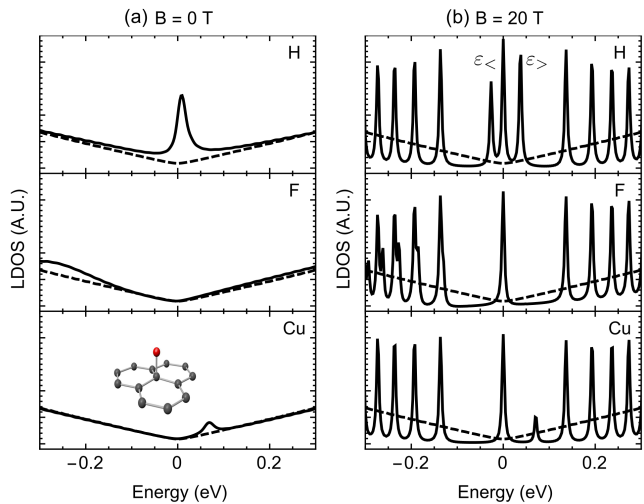}
\caption{LDOS of graphene with a H--, F--, and Cu--adatom in (a) ${B=0\,\text{T}}$, and (b) ${B=20\,\text{T}}$. 
LDOS is calculated numerically using TB model, at a carbon atom which is 8 and 10 unit cells away from the adatom, respectively. 
In order to enhance the resolution, the broadening is adjusted from 10~meV (${B=0\,\text{T}}$) to 3~meV (${B=20\,\text{T}}$).
LDOS of pristine graphene is shown for comparison (dashed line). 
When the magnetic field is absent, one can observe a single resonance peak either on electron or hole side depending on the on-site energy of the adatom. 
All the adsorbates are assumed to be in top position (inset) which is energetically favorable for the impurities considered \cite{Wehling2009}. 
All parameters used in the calculation are shown in Table~\ref{table:01}.
}
\label{fig:02}
\end{figure}

As shown in Fig.~\ref{fig:02}, we identify three distinct resonant behaviors depending on the kind of the impurity.

\paragraph{H--adatom.}
Hydrogen is known to act as a magnetic impurity and also to enhance spin-orbit coupling 
by corrugating graphene sheet \cite{Gmitra2013PRL, Irmer2015PRB, Kochan2017PRB}.
In this paper we focus on spin independent effects.
When the magnetic field is absent [see Fig.~\ref{fig:02} (a)], 
the resonance peak appears in the vicinity of the Dirac point ($\sim$7 meV) in the LDOS spectrum.
This resonance peak is formed by strong hybridization between the adatom and the carbon atom,
which is one of the main scattering mechanisms in transport \cite{Wehling2009, Wehling2010}.
In the presence of magnetic field [$B=20\,\text{T}$, see Fig.~\ref{fig:02}(b)], 
the resonance peak splits into two bound states ($\varepsilon_{<}=-27$ meV and $\varepsilon_{>}=37$ meV)
which are placed near the Dirac point between the zero and the first lowest LLs. The smaller the $\varepsilon_d$ is, the more symmetric
the bound state peaks would appear.

\paragraph{F--adatom.} 
The negative on-site energy of fluorine places the resonance peak deep in the valence band,
and strong hybridization with the states off the Dirac point 
makes the peak much broader.
First principles calculations also show that the adatom hybridizes strongly with the carbon atoms in graphene
to form mid-gap states in the valence band \cite{Irmer2018PRB}. 
When the external magnetic field is present, instead of
being the nearly symmetric,
those two bound state peaks are now shifted toward the valence band
without crossing LLs,
therefore, one peak is merged onto LL$_0$,
and the other on the right side of LL$_{-1}$.
Due to the broadening, the side peaks near the LLs are not pronounced, 
but small shoulders can still be seen in the valence band.

\paragraph{Cu--adatom.} 
Copper is not only a common impurity found, especially in CVD grown graphene,
but also an important functionalization element.
The resonance peak appears around 70 meV for $B=0\,\text{T}$, 
and unlike for H-- and F--adatoms, the peak position is almost inert to the external magnetic field [Fig.~\ref{fig:03}(b)].
The external magnetic field may slightly sharpen the peak, 
but there is no counterpart of the bound state in the valence band,
nor do any side peaks appear.

These three adatoms represent the three scenarios of the interaction between resonant impurity
and magnetic fields: (i) transition of one resonant state peak into nearly symmetric two bound states 
appearing around LL$_0$ due to the external magnetic field,
(ii) no pronounced bound states near the Dirac point but multiple side peaks next to LLs in one band, 
and (iii) a resonant state inert to the field. These different behaviors can be qualitatively understood
by investigating a limiting case, a single vacancy.

\begin{table}
\begin{tabular}{ C{1.8cm}||C{1.8cm}|C{1.8cm}|C{1.8cm}}
\hline
in eV & H & F & Cu\\ 
 \hline\hline
$\varepsilon_d$ & 0.16 & -2.2 & 0.08\\
$\omega_d$ & 7.5 & 5.5 & 0.81\\ 
\hline
\end{tabular}
\caption{Tight-binding parameters of adatoms. These values are fitted from the first principle calculations \cite{Irmer2015PRB, Frank2017PRB}.}
\label{table:01}
\end{table}

\subsection{Vacancy limit}
\begin{figure}[tbp]
\includegraphics[width=\columnwidth]{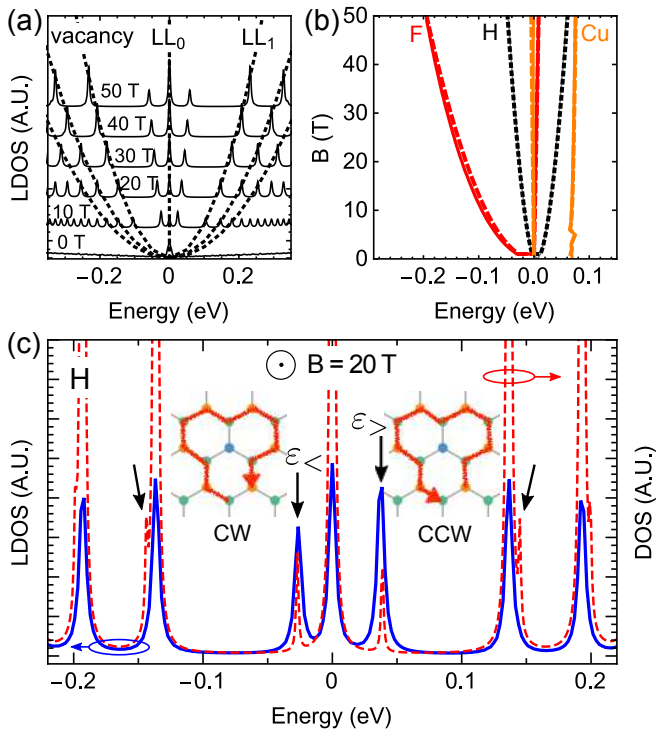}
\caption{
(a) Magnetic field dependence of LLs and bound state peaks due to vacancy.
Dashed lines represent LLs of pristine graphene as in Eq.~(\ref{eq:LLs}).
(b) The bound states energies due to different adatoms, H, F, and Cu at different magnetic fields. 
TB (solid) and Green's function (dashed) calculations agree with each other.
Near the zero energy, the separation between $LL_0$ and the bound states from F and Cu are smaller than the broadening.
(c) Comparison between LDOS (solid) from TB model and DOS (dashed) from Green's function for H--adatom when $B=20~\text{T}$.
The DOS units are scaled for the better comparison.
The two arrows in the center indicate the bound states ($\varepsilon_{<}$ and $\varepsilon_{>}$), and the other arrows indicate side peaks, which are the result of 
interactions between the H--adatom and higher LLs.
These side peaks are not visible in the TB calculation because of the broadening (3~meV).
The insets next to the bound state peaks are the schematics for the current density of the bound states.}
\label{fig:03}
\end{figure}

When a resonant impurity bonds to the underlying carbon atom, 
it effectively removes one $p_z$~orbital from the graphene.
In the limit of $\omega_d \to \infty$ and $\varepsilon_d\to0$,
it is equivalent to remove one carbon lattice site in our tight-binding model,
leaving a vacancy behind. This leaves the smallest zigzag edge around the vacancy,
so that strongly localized states can be present in the nearest neighbor sites in the vicinity of the vacancy.
Therefore, the vacancy is an idealized model for the resonant impurity with strong hybridization.
As in Eq.~(\ref{eq:LLs}), the magnetic field dependence of LLs proportional to $\sqrt{B}$ is shown in Fig.~\ref{fig:03}(a).
In addition to the LLs, with increasing $B$, 
the resonance peak splits into a pair of bound states  between LL$_0$ and LL$_{\pm1}$.
Interestingly, the bound state peaks appear to have the same $\sqrt{B}$ dependence as LLs.
and the bound states in the conduction band have their counterparts in the valence band at the same energy 
but with the opposite sign.

In Fig.~\ref{fig:03}(b), positions of the bound state peaks induced by adatoms H, F, and Cu are shown.
The asymmetry of the bound states in electron and hole side is due to $\varepsilon_{d}\neq0$.
For a given hybridization energy, the larger is the on-site energy of the adatom, the further away the peaks appear.
As mentioned above, the three possible patterns of bound states in external magnetic fields are 
nearly symmetric bound state peaks (H), 
a single (visible) bound state peak from F closely tracing LL$_{-1}$ (F), 
and a bound state peak which is qualitatively insensitive to the magnetic field (Cu).
Nonetheless, these three patterns are governed by the same mechanisms.

We first compare total density of states (DOS) calculated from the analytical Green's function method given by Eq.~(\ref{eq:DOS})
with the LDOS from the numerical TB model calculations in Fig.~\ref{fig:03}(c) for the graphene with H--adatom.
LDOS and DOS are rescaled for comparison, and the two results match closely.
The coinciding peak positions confirm the equivalence of the two complementary approaches.
The five tall peaks correspond to LL$_n$ ($|n| \leq 2$) and the two peaks near LL$_0$ are the bound states.
Each bound state has a current probability density circling around the impurity in the opposite direction.
This is schematically shown in the inset of Fig.~\ref{fig:03}(c).
The chiral local (probability or charge) current of the bound states generates an effective magnetic dipole,
and it lowers or increases the energy of the bound state depending on the chirality.
When an external magnetic field points out of the sheet, 
the probability current density of the lower (higher) energy bound state 
flows clockwise (counterclockwise) with the effective magnetic dipole moment of each state
aligned in the parallel (antiparallel) direction with respect to the magnetic field.
This explains the ordering of the bound states in the energy.
The impurity also interacts with higher LLs, 
and side peaks marked by arrows in Fig.~\ref{fig:03}(c) close to LL$_{\pm 1}$.
It is more transparent to consider the Green's function to analyze the relation between the two lowest bound states and the TB parameters.
The energies of the bound states can be analytically calculated by solving Eq.~(\ref{eq:approxGR}).
Focusing on the most pronounced bound state peaks, 
the energies of left and right lowest bound state peaks are $\varepsilon_{<}$ and $\varepsilon_{>}$.
As $\varepsilon_{d} > 0$ increases, the asymmetry of the peaks, represented by $|\varepsilon_{>}+\varepsilon_{<}|$,
also increases, and $\text{sgn}(\varepsilon_{d})=\text{sgn}(\varepsilon_{>}+\varepsilon_{<})$
while the strong hybridization $\omega_{d}\to \infty$ makes the system approach the vacancy limit (symmetric bound states).

One important conclusion derived from Eq.~(\ref{eq:approxGR}) is that 
the microscopic TB parameters, namely, $\varepsilon_{d}$ and $\omega_{d}$ 
can be determined from the lowest bound states energies:
\begin{eqnarray}\label{eq:Kochan-Lee formula}
\Bigg{\{}
\begin{matrix}
\varepsilon_{d}&\approx& (\varepsilon_{>}+\varepsilon_{<})\Big{/}
\left[ 1+2 H_{N_B} \frac{\varepsilon_{>}\varepsilon_{<}}{\varepsilon_{1}^2} \right]
\\
\omega_{d}^2&\approx&
- \frac{2}{\alpha}\frac{\varepsilon_{>}\varepsilon_{<}}{\varepsilon_{1}^2}\Big{/}
\left[ 1+2 H_{N_B} \frac{\varepsilon_{>}\varepsilon_{<}}{\varepsilon_{1}^2} \right]
\end{matrix},
\end{eqnarray}
where $\alpha=A_\text{uc}/2\pi (\hbar v_F)^2\approx0.0272$ and $\varepsilon_1=\hbar \omega_B$.
Equation~({\ref{eq:Kochan-Lee formula}}) 
provides a useful device that immediately links experimental measurements to microscopic TB parameters.
This formula can provide qualitatively reasonable estimates (within $\sim$ 10\% for broadening $0.1\,\text{meV}$) 
for $\varepsilon_d$ and $\omega_{d}$ as long as absolute values of the bound state energies $|\varepsilon_{>,<}|$ 
are greater than the resonance energy (for $B=0\,\text{T}$) 
but it should be noted that the accuracy reduces with increased broadening.
Therefore, Eq.~(\ref{eq:Kochan-Lee formula}) might not provide a reliable estimate
for some marginally resonant impurities, such as F, within a realistic range of the $B$--field strength.

\subsection{Charged impurity}
So far we considered neutral resonant impurities. 
However, charged impurities are also known to play an important role in transport not only as long-range scatterers
but also by forming electron-hole puddles \cite{Peres2010RMP}.
Recently, it was also demonstrated that the atomic collapse can occur due to local charged impurities \cite{Wang2013Science} 
or a vacancy \cite{Mao2016NPhys}.
Therefore, it is an interesting question to ask how a charge combined with a resonant impurity reacts to quantizing magnetic fields.
In order to investigate the effect of the charged impurity,
we add the screened Coulomb potential
\begin{eqnarray}\label{eq:coulomb}
U(r)=
\begin{cases}
-\hbar v_F \beta / r_0, \quad & r \leq r_0\\
-\hbar v_F \beta / r, \quad & r > r_0
\end{cases},
\end{eqnarray}
where $\beta=Z e^2/ \kappa \hbar v_F$ with the charge number $Z$ and the effective dielectric constant $\kappa$.
The Fermi velocity in graphene can be associated with the tight-binding hopping parameter as $\hbar v_F = 3/2 t a_0$.
A cutoff radius $r_0$ is introduced to prevent from diverging at $r=0$, and a realistic value $r_0=5\,\text{nm}$ is chosen 
to be consistent with experimental observations \cite{Wang2013Science}. 

\begin{figure}[tbp]
\includegraphics[width=\columnwidth]{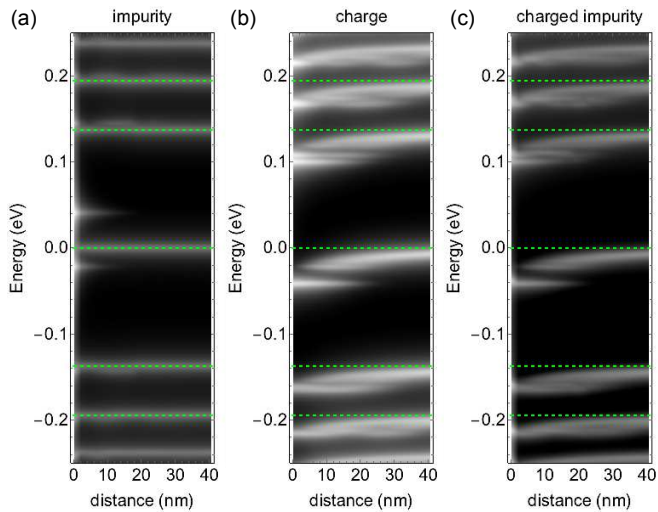}
\caption{
Spatial distribution of LDOS as a function of distance (along the armchair direction) 
from the scattering center (charged impurity) under the magnetic field $B=10\,\text{T}$. 
(a) The two bound states are pronounced in the close vicinity of the resonant impurity site contrasted with LLs that are 
evenly present over the distance.
(b) The LLs are bent down due to the positive Coulomb impurity.
(c) When the impurity is positively charged and resonant at the same time,
the Coulomb interaction dominates. 
The broadening is 3 meV, and $\beta= 0.357$, $t=3.7\,\text{eV}$, and the distance between the charge 
and the graphene sheet $d=0.6\,\text{nm}$ are used.
\label{fig:04}
}
\end{figure}

Experimentally, the energies of the bound states can be measured by scanning tunneling spectroscopy (STS) 
as differential conductivity $dI/dV$, which can be directly mapped on to the LDOS spectrum.
In Fig.~\ref{fig:04} we show calculated LDOS as a function of the distance from the impurity site (armchair direction)
for $B=10\,\text{T}$. The brighter shade represents the greater values of LDOS.
The magnetic field induced LLs appear as the horizontal lines over a distance
while the bound states formed by the resonant impurity (H--adatom) are pronounced only in the vicinity of 
the impurity [Fig.~\ref{fig:04}(a)]. 
A positively charged impurity in the substrate lowers the energy and the LLs bent down 
around the impurity sites [Fig.~\ref{fig:04}(b)].
This result demonstrates the splitting of the LLs due to the orbital degeneracy lifting
as in the experimental observation near a single charged vacancy \cite{Luican-Mayer2014PRL}.
In case that the resonant impurity is also positively charged, or an adatom is located on top of a charged impurity in the substrate,
the bound state is not very clearly distinguished, 
and the Coulomb interaction supersedes the resonant scattering [Fig.~\ref{fig:04}(c)].
The strong resonant impurity exhibits qualitatively the same behavior as vacancy 
and this implies that the Coulomb interaction provides greater contribution to the scattering amplitude
under the magnetic field.

\begin{figure}[tbp]
\includegraphics[width=\columnwidth]{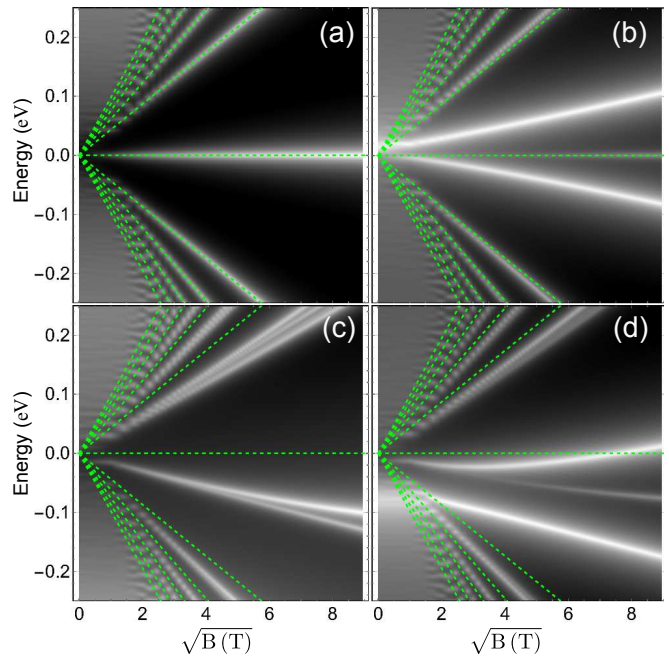}
\caption{
Landau fan diagram of graphene with (a) a resonant impurity, (b) a charged impurity, 
(c) a charged resonant impurity at a distance $d=0\,\text{nm}$,
and (d) a charged resonant impurity at a distance $d=2.6\,\text{nm}$.
The dashed lines indicate the Landau levels of pristine graphene.
The dashed lines indicate LLs in pristine graphene.
The charge $\beta=0.4$ is used. 
\label{fig:05}
}
\end{figure}

When a stronger magnetic field is applied, resonance-induced bound states can still reappear.
Figure \ref{fig:05}(a) shows the fan-diagram of pristine graphene within a range of magnetic fields.
With an H--adatom, the two bound states separated by $LL_0$ are distinctly seen [Fig.~\ref{fig:05}(b)]
same as in Figs.~\ref{fig:03}(c) and \ref{fig:04}(a).
The energy splitting between the two levels are proportional to $\sqrt{B}$.
A positively charged resonant impurity [Fig.~\ref{fig:05}(c)] shifts all the LLs to lower energies, 
and also splits LL$_1$ into two orbital states. 
If the charge becomes greater than the critical charge $\beta_c = 0.5$, 
then the lower orbital state from LL$_1$ evolves into an atomic collapse resonance state \cite{Moldovan2016Thesis}.
As seen in Fig.~4, in the simultaneous presence of resonance and charge,
charge scatters stronger than the resonance, and the trace of resonance is not so pronounced. But in higher magnetic fields, as shown in Fig.~\ref{fig:05}(d), 
the bound state peak splits out of the shifted LL$_0$ with a higher LDOS intensity.

\section{Conclusion}
We have performed realistic calculations of the electronic properties 
of adsorbates on graphene in the presence of a transverse quantizing magnetic field.
Three adatoms were investigated as special examples: H, F, and Cu, each with distinct
binding characteristics. The interaction between resonant adatoms and strong magnetic fields
leads to specific bound states around the Dirac point with unique spectral features
which can be explored experimentally. In particular, such features could be used to 
extract useful microscopic model parameters of the resonant adsorbate.
In principle,
the same framework can be applied to investigating the magnetic exchange,
local spin-orbit coupling, or other spin dependent effects in the means of 
STS measurements even without requiring the spin resolution.
We have also compared adatoms with long-range scatterers. In a low magnetic field, 
the Coulomb impurity is more pronounced, and a high magnetic field
activates the resonance-induced bound states.

\begin{acknowledgments}
This work was supported by the DFG 1277 (A09) and 
the EU Horizon 2020 Framework Programme under Grant Agreement 696656.
\end{acknowledgments}


\begin{thebibliography}{}
\bibitem{CastroNeto2009RMP}
A.~H. Castro~Neto, F. Guinea, N.~M.~R. Peres, 
K.~S. Novoselov, and A.~K. Geim, Rev. Mod. Phys. {\bf 81}, 109 (2009).

\bibitem{Wehling2009} 
T.~O. Wehling, M.~I. Katsnelson, and, A.~I. Lichtenstein, 
Chem. Phys. Lett. {\bf 476}, 125 (2009).

\bibitem{Wehling2010} 
T.~O. Wehling, S. Yuan, A.~I. Lichtenstein, A.~K. Geim, and M.~I. Katsnelson,
Phys. Rev. Lett. {\bf 105}, 056802 (2010).

\bibitem{Loktev2007PRB}
Y.~V. Skrypnyk and V.~M. Loktev
Phys. Rev. B {\bf 75}, 245401 (2007).

\bibitem{Loktev2010PRB}
Y.~V. Skrypnyk and V.~M. Loktev,
Phys. Rev. B {\bf 82}, 085436 (2010).

\bibitem{Han2014NNano}
W. Han, R.~K. Kawakami, M. Gmitra, and J. Fabian, 
Nature Nanotechnol. {\bf 9}, 794 (2014).

\bibitem{CastroNeto2009PRL} 
A.~H. Castro Neto, F. Guinea, 
Phys. Rev. Lett. {\bf 103}, 026804 (2009).

\bibitem{Weeks2011PRX}
C. Weeks, J. Hu, J. Alicea, M. Franz, and R. Wu,
Phys. Rev. X {\bf 1}, 021001 (2011).

\bibitem{Mertig2010PRB}
S. Abdelouahed, A. Ernst, J. Henk, I.~V. Maznichenko, and I. Mertig,
Phys. Rev. B {\bf 82}, 125424 (2010).

\bibitem{Zhou2010Carbon}
J. Zhou, Q. Liang, and J. Dong,
Carbon {\bf 48}, 1405 (2010).

\bibitem{Avsar2015TDMatt}
A. Avsar, J.~H. Lee, G.~K.~W. Koon, and B. {\"O}zyilmaz,
2D Mater. {\bf 2}, 044009 (2015).

\bibitem{Balakrishnan2013NPhys}
J. Balakrishnan, G.~K.~W. Koon, M. Jaiswal, A.~H. Castro Neto, 
and B. Özyilmaz,
Nature Phys. {\bf 9}, 284 (2013).

\bibitem{Pachoud2014PRB}
A. Pachoud, A. Ferreira, B.~{\"O}zyilmaz, and A.~H. Castro Neto,
Phys. Rev. B {\bf 90}, 035444 (2014).

\bibitem{Zollner2016PRB} 
K. Zollner, T. Frank, S. Irmer, M. Gmitra, D. Kochan, and J. Fabian,
Phys. Rev. B {\bf 93}, 045423 (2016).

\bibitem{Gmitra2013PRL} 
M. Gmitra, D. Kochan, and J. Fabian,
Phys. Rev. Lett. {\bf 110}, 246602 (2013).

\bibitem{Kochan2014PRL} 
D. Kochan, M. Gmitra, and J. Fabian,
Phys. Rev. Lett. {\bf 112}, 116602 (2014).

\bibitem{Frank2017PRB} 
T. Frank, S. Irmer, M. Gmitra, D. Kochan, and J. Fabian,
Phys. Rev. B {\bf 95}, 035402 (2017).

\bibitem{Kochan2017PRB} 
D. Kochan, S. Irmer, and J. Fabian,
Phys. Rev. B {\bf 95}, 165415 (2017).

\bibitem{Yazyev2007PRB}
O.~V. Yazyev and L. Helm,
Phys. Rev. B {\bf 75}, 125408 (2007).

\bibitem{Hong2011PRB}
X. Hong, S.-H. Cheng, C. Herding, and J. Zhu,
Phys. Rev. B {\bf 83}, 085410 (2011).

\bibitem{McCreary2012PRL}
K.~M. McCreary, A.~G. Swartz, W. Han, J. Fabian, and R.~K. Kawakami,
Phys. Rev. Lett. {\bf 109}, 186604 (2012).

\bibitem{Gonzalez2016Science}
H. Gonz{\'a}lez-Herrero, J.~M. G{\'o}mez-Rodr{\'i}guez, P. Mallet,
M. Moaied, J.~J. Palacios, C. Salgado, M.~M. Ugeda, J.-Y. Veuillen, 
F. Yndurain, and I. Brihuega,
Science {\bf 352}, 438 (2016).

\bibitem{Tang2105SR}
T. Tang, N. Tang, Y. Zheng, X. Wan, Y. Liu, F. Liu, Q. Xu, and Y. Du,
Sci. Rep. {\bf 5}, 8448 (2015).

\bibitem{Geim2010NLett}
Z.~H. Ni, L.~A. Ponomarenko, R.~R. Nair, R. Yang, S. Anissimova,
I.~V. Grigorieva, F. Schedin, P. Blake, Z.~X. Shen,
E.~H. Hill, K.~S. Novoselov, and A.~K. Geim,
Nano Lett. {\bf 10}, 3868 (2010).

\bibitem{Ozyilmaz2011ACSNano}
M. Jaiswal, C.~H.~Y.~X. Lim, Q. Bao, C.~T. Toh, K.~P. Loh,
and B. {\"O}zyilmaz,
ACS Nano {\bf 5}, 888 (2011).

\bibitem{Zhu2015PRB}
A.~A. Stabile, A. Ferreira, J. Li, N.~M.~R. Peres, and J. Zhu,
Phys. Rev. B {\bf 92}, 121411(R) (2015).

\bibitem{Kawakami2018PRL}
J. Katoch, T. Zhu, D. Kochan, S. Singh, J. Fabian, and R.~K. Kawakami,
Phys. Rev. Lett. {\bf 121}, 136801 (2018).

\bibitem{Andrei2012RPP}
E.~Y. Andrei, G. Li, and X. Du,
Rep. Prog. Phys. {\bf 75}, 056501 (2012).

\bibitem{Silvestrov2014PRB} 
P.~G. Silvestrov,
Phys. Rev. B {\bf 90}, 235130 (2014).

\bibitem{Kaverzin2015PRB}
A.~A. Kaverzin and B.~J. van Wees
Phys. Rev. B {\bf 91}, 165412 (2015).

\bibitem{Wojtaszek2011JAP}
M. Wojtaszek, N. Tombros, A. Caretta, P.~H.~M. van Loosdrecht, and B.~J. van Wees
J. Appl. Phys. {\bf 110}, 063715 (2011).

\bibitem{Irmer2018PRB}
S. Irmer, D. Kochan, J. Lee, and J. Fabian,
Phys. Rev. B {\bf 97}, 075417 (2018)

\bibitem{Irmer2015PRB} 
S. Irmer, T. Frank, S. Putz, M. Gmitra, D. Kochan, J. Fabian,
Phys. Rev. B {\bf 91}, 115141 (2015).

\bibitem{Sarma2011RMP}
S. Das Sarma, S. Adam, E.~H. Hwang, and E. Rossi
Rev. Mod. Phys. {\bf 83}, 407 (2011).

\bibitem{Luican-Mayer2014PRL}
A. Luican-Mayer, M. Kharitonov, G. Li, C.-P. Li, I. Skachko, 
A.-M.~B. Gon\c{c}alves, K. Watanabe, T. Taniguchi, and E.~Y. Andrei, 
Phys. Rev. Lett. {\bf 112}, 036804 (2014).

\bibitem{Andrei2016NPhys}
J. Mao, Y. Jiang, D. Moldovan, G. Li, K. Watanabe, T. Taniguchi, 
M.~R. Masir, F.~M. Peeters, and E.~Y. Andrei,
Nature Phys. {\bf 12}, 545 (2016).

\bibitem{footnote:01}
$3.7\,\text{eV}$ is used for Figs. \ref{fig:04} and \ref{fig:05} 
in order to fit the experimental data in Ref.~\cite{Luican-Mayer2014PRL}.

\bibitem{footnote:02}
A square flake with edge length $\sim$162~nm (Figs. \ref{fig:01}-\ref{fig:03}), 
and a hexagon flake with edge length $\sim$100~nm (Figs. \ref{fig:04}-\ref{fig:05}) are considered.
The square flake consists of two Zigzag edges and two armchair edges, 
while the hexagon flake has only armchair edges, 
however, the geometries or the edges do not alter our results.

\bibitem{Rusin2011} 
T.~M. Rusin, and W. Zawadzki,
J. Phys. A: Math. Theor. {\bf 44} 105201 (2011).

\bibitem{Weisse2006RMP}
A. Wei{\ss}e, G. Wellein, A. Alvermann, and H. Fehske,
Rev. Mod. Phys. {\bf 78}, 275 (2006).

\bibitem{Moldovan2016}
D. Moldovan and F. M. Peeters, \textsc{pybinding}: A
  \textsc{python} package for tight-binding calculations,
http://dx.doi.org/10.5281/zenodo.56818.

\bibitem{Loewdin1951}
P.-O. L{\" o}wdin,
J. Chem. Phys. {\bf 19}, 1396 (1951).

\bibitem{Callaway1964}
J. Callaway, 
J. Math. Phys. {\bf 5}, 783 (1964).

\bibitem{Zheng2002PRB}
Y. Zheng and T. Ando,
Phys. Rev. B {\bf 65}, 245420 (2002).

\bibitem{Peres2006PRB}
N.~M.~R. Peres, F. Guinea, and A.~H. Castro~Neto,
Phys. Rev. B {\bf 73}, 125411 (2006).

\bibitem{Horing2010}
N.~J.~ M. Horing,
Phil. Trans. R. Soc. A {\bf 368}, 5525 (2010).

\bibitem{Nanda2012NJP}
B.~R.~K. Nanda, M. Sherafati, Z.~S. Popovi{\' c}, and S. Satpathy, 
New J. Phys. {\bf 14}, 083004 (2012).

\bibitem{Bundesmann2015PRB} 
J. Bundesmann, D. Kochan, F. Tkatschenko, J. Fabian, and K. Richter,
Phys. Rev. B {\bf 92}, 081403(R) (2015).

\bibitem{Peres2010RMP}
N.~M.~R. Peres,
Rev. Mod. Phys. {\bf 82}, 109 (2010).

\bibitem{Wang2013Science} 
Y. Wang, D. Wong, A.~V. Shytov, V.~W. Brar, S. Choi, Q. Wu, H.-Z. Tsai, W. Regan, 
A. Zettl, R.~K. Kawakami, S.~G. Louie, L.~S. Levitov, and M.~F. Crommie,
Science {\bf 340}, 734 Science (2013).

\bibitem{Mao2016NPhys}
J. Mao, Y. Jiang, D. Moldovan, G. Li, K. Watanabe, T. Taniguchi, M.~R. Masir, F.~M. Peeters, and E.~Y. Andrei,
Nature Phys. {\bf 12}, 545 (2016).

\bibitem{Moldovan2016Thesis}
D. Moldovan, Ph.D. thesis, University of Antwerp, 1980.

\end{thebibliography}
\end{document}